\documentclass[fleqn,usenatbib]{mnras}

\usepackage{newtxtext,newtxmath}
\usepackage[T1]{fontenc}

\DeclareRobustCommand{\VAN}[3]{#2}
\let\VANthebibliography\thebibliography
\def\thebibliography{\DeclareRobustCommand{\VAN}[3]{##3}\VANthebibliography}

\usepackage{graphicx}	% Including figure files
\usepackage{amsmath}	% Advanced maths commands

\newcommand{\BB}{{\bf {B}}}

\newcommand{\xx}{{\bf {x}}}
\newcommand{\yy}{{\bf {y}}}

\newcommand{\rr}{{\bf {r}}}

\newcommand{\half}{\textstyle{\frac{1}{2}}}

\newcommand*\id{\mathop{}\!\mathrm{d}}

\title[Solar atmospheric heating at a fin current sheet ]{Heating in the Solar Atmosphere at a Fin Current Sheet Driven by Magnetic Flux Cancellation}

\author[E.R. Priest and D. I. Pontin]{
Eric R. Priest,$^{1}$\thanks{E-mail: eric.r.priest@gmail.com}
David I. Pontin$^{2}$\thanks{E-mail: david.pontin@newcastle.edu.au}
\\
$^{1}$School of Mathematics and Statistics, 
University of St Andrews, 
St Andrews, KY16 9SS, UK\\
$^{2}$School of Information and Physical Sciences, University of Newcastle, Callaghan, NSW 2308, Australia
}

\date{Accepted XXX. Received YYY; in original form ZZZ}

\pubyear{\the\year{}}

\begin{document}
\label{firstpage}
\pagerange{\pageref{firstpage}--\pageref{lastpage}}
\maketitle

% Abstract of the paper
\begin{abstract}
Magnetic reconnection before flux cancellation in the solar photosphere when two opposite-polarity photospheric magnetic fragments are approaching one another is usually modelled by assuming that a small so-called ``floating current sheet" forms about a null point or separator that is situated in the overlying atmosphere. Here instead we consider the reconnection that is initiated as soon as the fragments become close enough that their magnetic fields interact. The resulting current sheet, which we term a ``fin sheet" extends up from the null point or separator that is initially located in the solar surface. We  develop here nonlinear analyses for finite-length models of both fin and  floating current sheets that extend the previous models that were limited to short floating current sheets.  These enable the length of the current sheet to be calculated in both cases as functions of the separation distance of the sources and the reconnection rate, as well as the rate of heating.  Usually, the fin current sheet liberates more energy than a floating current sheet.
\end{abstract}

% Select between one and six entries from the list of approved keywords.
% Don't make up new ones.

%%%%%%%%%%%%%%%%%%%%%%%%%%%%%%%%%%%%%%%%%%%%%%%%%%

%%%%%%%%%%%%%%%%% BODY OF PAPER %%%%%%%%%%%%%%%%%%

\section{Introduction} \label{sec:intro}

\cite{pontin24} have proposed that magnetic reconnection driven by photospheric flux cancellation may contribute substantially to the heating of the solar corona and the acceleration of the solar wind. Such reconnection has for long been associated with a range of dynamic events in the solar atmosphere, such as X-ray  right points \citep{martin85,parnell94a,priest94,parnell95,archontis14}, X-ray jets \citep{shibata92,shimojo07}, UV bursts in active regions \citep{peter14}, and transition-region explosive events \citep{brueckner83,innes97}. Flux cancellation was thought before 2010 to take place mainly near the boundaries of supergranules, since that was where most of the  photospheric magnetic flux observed at that time was concentrated \citep[e.g.,][]{schrijver97a,priest14}.

However, ground-breaking observations from the SUNRISE balloon mission \citep{lagg10,solanki11,solanki17b,smitha17} have transformed our view by showing that flux cancellation is very much more common than previously thought and often occurs around granules. This has been shown to imply that flux cancellation may be a viable mechanism for heating the chromosphere and corona \citep{priest18}, and a series of models for the process has been developed \citep{priest21a,syntelis19a,syntelis20,syntelis21}. In addition, clear examples of
%brightenings 
impulsively enhanced emission
triggered by flux cancellation have been discovered, including in coronal loops \citep{tiwari14,huang18}, in association with UV bursts and birectional jets \citep{chitta17a} and campfires \citep{panesar21}, in the cores of active regions \citep{chitta18,chitta20}, and either between bipolar footpoints or between such a footpoint and an opposite-polarity photospheric magnetic fragment \citep{kahil22}.

In response to slow motions of the photospheric footpoints of coronal magnetic fields on time-scales longer than the Alfv\'en travel time, the magnetic field evolves through a series of magnetohydrostatic equilibria. If the plasma pressure is much smaller than the magnetic pressure and the height of the structure is less than the gravitational scale-height, those equilibria will be force-free in nature, and, if the electric currents are small, they will be potential magnetic fields. When the equilibria are topologically complex, in the sense that they possess null points or separator field-lines, then current sheets can form around such nulls or separators and magnetic reconnection can take place at them.  Current sheets can also form in topologically simple but geometrically complex configurations where nulls and separators are absent but quasi-nulls or quasi-separators are present \citep[e.g.,][and references therein]{pontin22}.

There are many ways in which current sheets can appear and evolve, as described in \cite{priest14}. Several techniques have been employed for modelling current sheets, each of which have pros and cons \citep{priest14}.  For modelling the appearance and slow evolution of current sheets in two dimensions, an invaluable and elegant technique is to use complex variable theory, as first proposed by \cite{green65} and \cite{tur76}. Modelling them in three dimensions is much more difficult, since complex variable theory is no longer valid, but a method has been developed by \cite{priest21a} for an axisymmetric three-dimensional sheet by treating the sheet as a series of current rings \citep[see also][]{longcope96a,longcope96b}. For non-axisymmetric fields or to describe the time-dependent fragmentation of sheets into plasmoids or flux ropes, it is generally necessary instead to use computational MHD modelling. In this paper, on grounds of simplicity and ease of determination of the response to varying the reconnection rate or the footpoint locations, we model current sheets as two-dimensional structures in an ambient potential field.

In general, several types of models  may be constructed that complement one another and give different insights about the basic physics of complex solar phenomena. On the one hand, it is possible to develop complex three-dimensional computational models that include as much realistic physics as possible. These can produce images or spectra that may be compared with detailed observations and have led to great advances in  interpreting the observations. However, they are often extremely difficult to understand and are unable to "simulate'' fully the observations, since there is always some physics that is omitted and they cannot achieve the spatial resolution that is necessary for a full understanding. In addition, they can be run for only a few specific ranges of parameters and for limited time, due to computational constraints.  Here we adopt an alternative philosophy of constructing much simpler, largely analytical models, in which one can easily see the effects of a wide range of parameters.  They use in-depth physical understanding to focus on key physical processes (here magnetohydrodynamic) while omitting others, but they are also able to produce key advances in understanding in order to interpret observations and to guide future computational experiments.

When two photospheric magnetic flux patches of opposite polarity approach one another, reconnection may be initiated at a current sheet between them. Previous modelling has assumed the current sheet forms about a null point or separator that is situated above the solar surface, typically in the corona, so that the current sheet is also  detached from the surface, a situation that we refer to as a ``floating" current sheet.
However, there is significant evidence that the reconnection may in fact occur lower in the atmosphere. Transient Ca II H brightenings observed by \emph{Hinode} \citep{park2012} suggest that reconnection associated with flux cancellation may be happening in the chromosphere or even in the photosphere \citep{litvinenko2015}. \cite{nelson2013} present both observations and simulations of Ellerman Bombs initiated when flux cancellation drives magnetic reconnection at chromospheric heights. \cite{shelyag2018} analysed simulations of magnetoconvection, identifying numerous discrete flux cancellation events in which the associated magnetic reconnection occurred in the photospheric layer in a so-called flux pile-up regime.

We here consider an alternative scenario to that of a floating current sheet, in which the current sheet grows up from the photosphere as the flux patches approach, and  we call this a ``fin'' current sheet due to its resemblance to the dorsal fin of a sea mammal (Section \ref{sec:2d}). Related ideas were considered briefly by \cite{low88a,low91c}. Such initially low-lying sheets may be important when energy is released low in the atmosphere, as, for example in Ellerman bombs in the low chromosphere \citep{rouppe16,hansteen17}. 
In addition, we generalise the previous modelling of a floating current sheet by dropping the previous assumption that the current sheet be ``small" in the sense that its vertical dimension is much smaller than the typical distance between flux sources (see Section \ref{sec:float}). 

Both floating and fin current sheets are likely to be important in the solar atmosphere, but their presence depends on the nature of both the magnetic topology and the motion of the phospheric magnetic fragments which can be regarded as acting as sources for coronal magnetic fields. If the topology is relatively simple and one starts with two isolated magnetic fragments that are not connected magnetically, then, as they approach one another, their magnetic fields will come into contact and a fin current sheet will naturally form.  On the other hand, if the flux sources are initially connected magnetically and they start to move towards one another, then their motion will drive the formation of a floating current sheet along the separator or around the null point that was initially present. Considering a region whose topology is complex with null points or separators being present,  the motion of photospheric footpoints will naturally form floating current sheets.

The paper is organised as follows. In Section \ref{sec:2d} we present a new model for reconnection at a fin current sheet. Then in Section \ref{sec:float} we extend previous models of a floating current sheet to finite length sheet. Finally, in Section \ref{sec:conc} we present a discussion, including a comparison of the energy release in the two cases.

\section{Fin current sheet model}\label{sec:2d}
\subsection{Magnetic Field Model}
\label{sec2.1}

Consider a two-dimensional magnetic field ($B_x,B_y$) in the solar corona (represented by the half-space $y>0$) generated by two  patches of magnetic flux in the presence of a uniform background horizontal field ($B_0\hat{\bf x}$) of strength $B_0$. For simplicity we model the flux patches (which represent magnetic flux that is generated in the solar interior and is threading through the solar surface into the corona) as point sources located at $x=\pm d$ on the $y$-axis. 
\begin{figure*}
\centering
\includegraphics[width=0.99\textwidth]{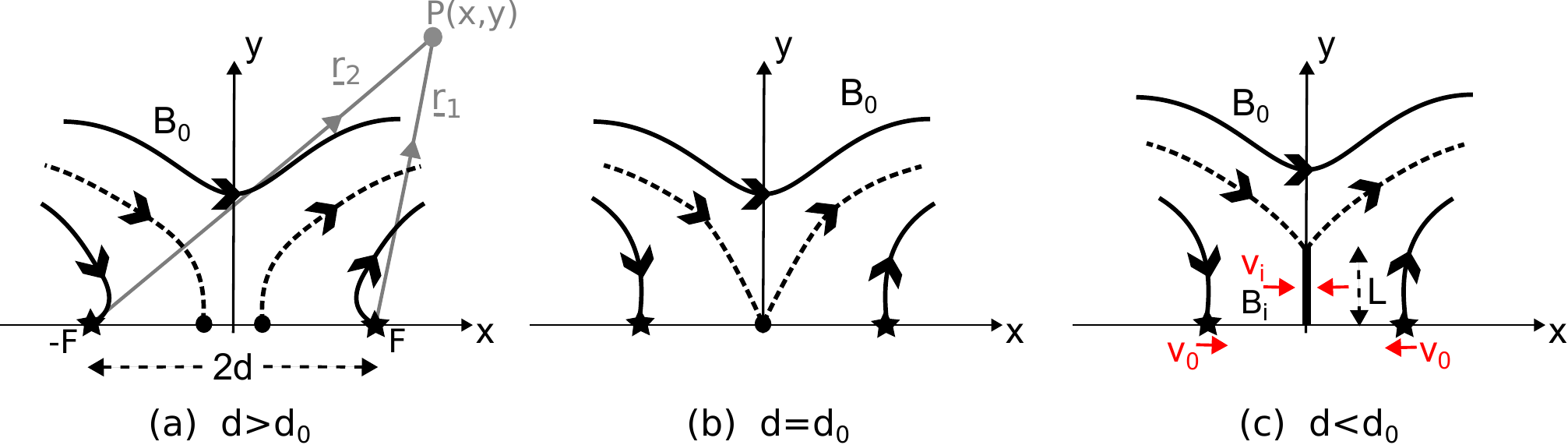}
\caption{Two magnetic flux sources of flux $\pm F$ are situated on the $x$-axis  a distance $2d$ apart in the presence of an overlying uniform magnetic field $B_0$. The distances from the sources to a point P($x,y$) are given by vectors ${\bf r}_1$ and ${\bf r}_2$. As $d$ decreases and the sources approach one another, there are several stages: (a) when $d>d_0=2F/(\pi B_0)$, there are two linear null points (large dots) on the $x$-axis; (b) when $d=d_0$, the nulls combine to form a second-order null; (c) when $d<d_0$, a current sheet forms of length $L$ at which the inflowing magnetic field is $B_i$ and the inflow speed is $v_i$.} 
\label{fig:sketch}
\end{figure*}
We denote the half-separation of the sources as $d$ (see Fig.~\ref{fig:sketch}) and consider what happens as $d$ decreases so that the flux patches approach one another. Here we suppose that the time-scale for approach is much longer than the Alfv\'en travel time, so that the approach speed is much smaller than the Alfv\'en speed, and the evolution is quasi-static through a series of potential fields.
The resulting potential magnetic field is given by 
\begin{equation}\label{eq:Bfield}
    \BB=\frac{F\hat{\rr}_1}{\pi |\rr_1|} - \frac{F\hat{\rr}_2}{\pi |\rr_2|} + B_0\,\hat{\bf x},
\end{equation}
where 
\begin{equation}
    \rr_{1,2}=(x\pm d)\,\hat{\xx}+y\hat{\yy}
    \nonumber
\end{equation} 
are the vector positions of the point ($x,y$) relative to the two sources and $\pm F$ are the strengths of the sources.

The so-called  ``flux interaction distance" \citep{longcope98} is given by
\begin{equation}
    d_0 \equiv \frac{2F}{\pi B_0}.
\end{equation}
When the sources are sufficiently far apart ($d>d_0$), the two sources are not joined by any magnetic field lines, but two magnetic null points are located between them on the $x$-axis (Fig.~\ref{fig:sketch}a).
%As the sources approach one another, $d$ decreases and eventually the 
As the parameter $d$ is decreased, the two null points approach one another, and when $d=d_0$ they will
coincide at the origin
(Fig.~\ref{fig:sketch}b). 
For the magnetic field described in Eqn.(\ref{eq:Bfield}), 
%We non-dimensionalise the problem using this length-scale, and write $\bar{d}=d/d_0$, $\bar{\rr}=\rr/d_0$, as well as $\bar{\BB}=\BB/B_0$. With this non-dimensionalisation 
the $x$-component of the magnetic field on the $y$-axis is
\begin{equation}
    B_x=B_0\left( 1-\frac{d_0d}{y^2+d^2}\right),
    \nonumber
\end{equation}
and so, when $d<d_0$, there is a null point present on the positive $y$-axis, located at $y_N=\sqrt{d_0d-d^2}$. 

Now, consider what happens as the sources approach one another, starting from $d>d_0$ and ending up with $d<d_0$. In a vacuum, the magnetic field would simply pass through a series of potential fields as the two null points on the $x$-axis approach one another, merge, and then a null point lifts off the surface along the $y$-axis, while the fields continuously change their topology.  However, in an ideal plasma, changes of topology are not allowed and so a current sheet would form with a length determined by magnetic flux conservation, as shown in Fig. \ref{fig:sketch}c.  Again, in a non-ideal plasma with, for example, finite resistivity, the reconnection rate would determine the changes in magnetic fluxes and therefore the size of the current sheet, as follows.

To represent the idealised magnetic field containing a tangential discontinuity we follow  \cite{green65,syrovatsky71,tur76} in using a complex-variable representation with $z=x+iy$. We can write the magnetic field (\ref{eq:Bfield}) as
\begin{align}
   B_x-iB_y &= B_0\left(1+\frac{d_0}{2(z-d)} - \frac{d_0}{2(z+d)}\right)
   \nonumber\\
   &= B_0\frac{z^2-d^2+d_0d}{z^2-d^2}.
\end{align}
When $d=d_0$ this reduces to $B_x-iB_y = B_0z^2/(z^2-d_0^2)$, and so for $d<d_0$ we choose a magnetic field that possesses flux sources at $z=\pm d$, tends to $B_x=B_0$ at infinity,  includes a branch cut from $z=iL$ to $z=-iL$ to represent the current sheet, and reduces to the same form when $L=0$, namely,
\begin{equation}\label{eq:Bcut}
    B_x-iB_y = B_0\frac{z(z^2+L^2)^{1/2}}{z^2-d^2}.
\end{equation}
For the upper half plane ($z\geq 0$), this has a fin current sheet (a cut in the positive complex plane ($y\geq 0$)) from $y=0$ to $y=L$.

\subsection{Fin Current Sheet Length (L(d)) with no Reconnection}
\label{sec2.2}
The length of the current sheet ($L$) is expected to be a function of the source separation ($2d$). It is determined by a constraint that comes from conservation of the flux above the current sheet. Consider first the case where reconnection is prohibited. In that case the flux above the current sheet must remain fixed as the flux patches approach one another. In order to calculate this flux we note that, for $d<d_0$, Eqn.(\ref{eq:Bcut}) gives the following expressions for the field components on the $y$-axis
\begin{align}
    B_x &= B_0\left\{ 
    \begin{array}{cc}
        0 & y<L \\
        y(y^2-L^2)^{1/2}/(y^2+d^2) & y>L
    \end{array}
    \right. \nonumber\\
    B_y &= B_0\left\{
    \begin{array}{cc}
        \pm y(L^2-y^2)^{1/2}/(y^2+d^2) & y<L \\
        0 & y>L. \\
    \end{array}
    \right. \label{eq:by_2d}
\end{align}
Thus, equating the flux across the $y$-axis between $y=L$ and $y=\infty$ when $d<d_0$ to the corresponding flux between $y=0$ and $y=\infty$  when $d=d_0$ yields
\begin{equation}
    \int_L^\infty \frac{y(y^2-L^2)^{1/2}}{y^2+d^2} \id y 
    = \int_0^\infty \frac{y^2}{y^2+d_0^2}\id y.
    \nonumber
\end{equation}
Since both of these integrals are infinite due to the presence of the uniform component $B_0\,\hat{\bf x}$ in Eqn.(\ref{eq:Bfield}), we subtract the integral over the uniform field out to infinity from both sides to give
\begin{align*}
%\begin{equation}
    L + \int_L^\infty 1-\frac{y(y^2-L^2)^{1/2}}{y^2+d^2} \id y 
    = \int_0^\infty 1-\frac{y^2}{y^2+d_0^2}\id y,\nonumber
    \nonumber
%\end{equation}
\end{align*}
where the right-hand side is simply
\begin{equation}
    \int_0^\infty \frac{d_0^2}{y^2+d_0^2}\id y
    =\left[d_0 \arctan \frac{y}{d_0}\right]_0^\infty
    =\frac{\pi}{2}d_0.
    \nonumber
\end{equation}

\begin{figure}
\centering
(a)\includegraphics[width=0.4\textwidth]{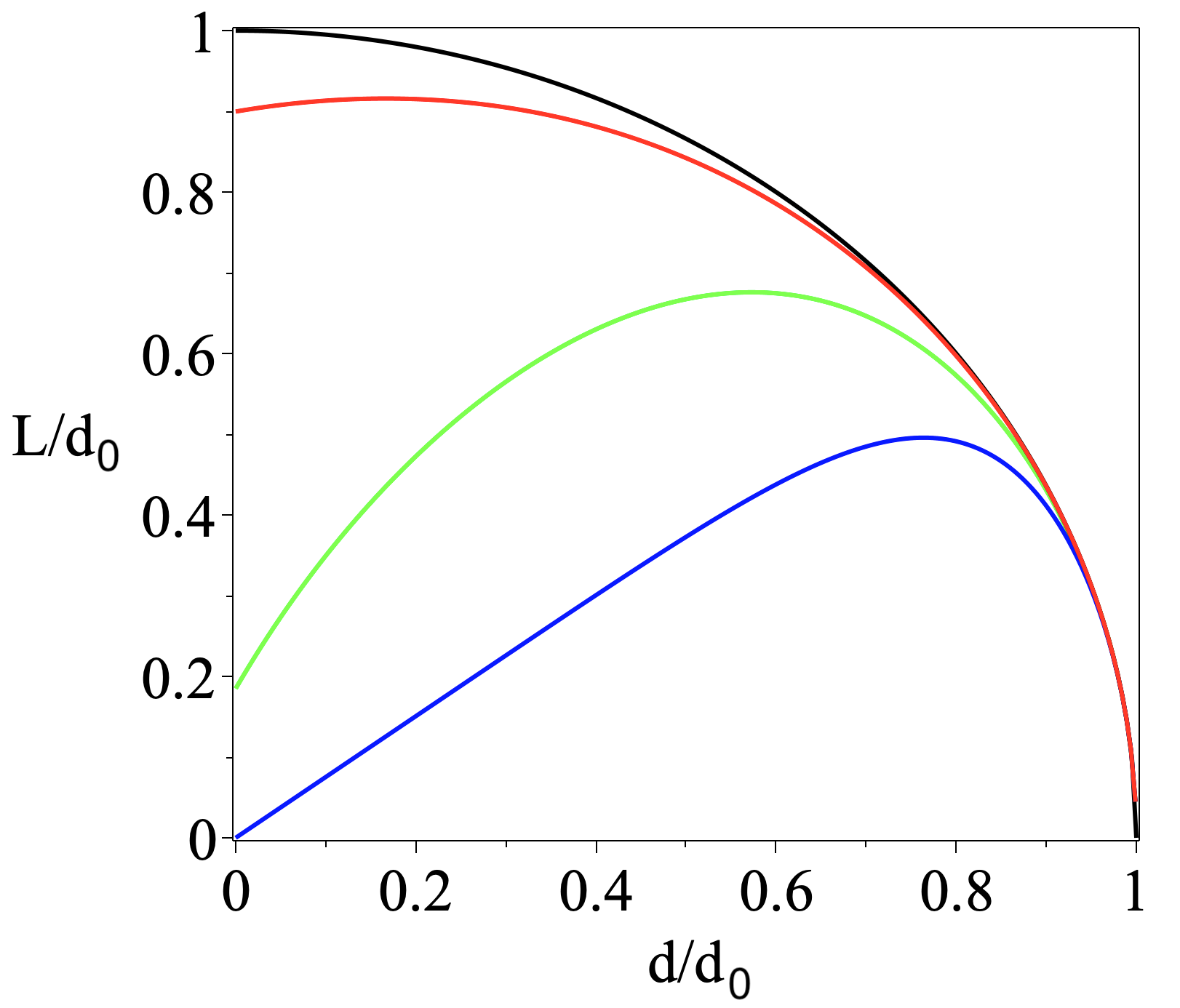}
(b)\includegraphics[width=0.4\textwidth]{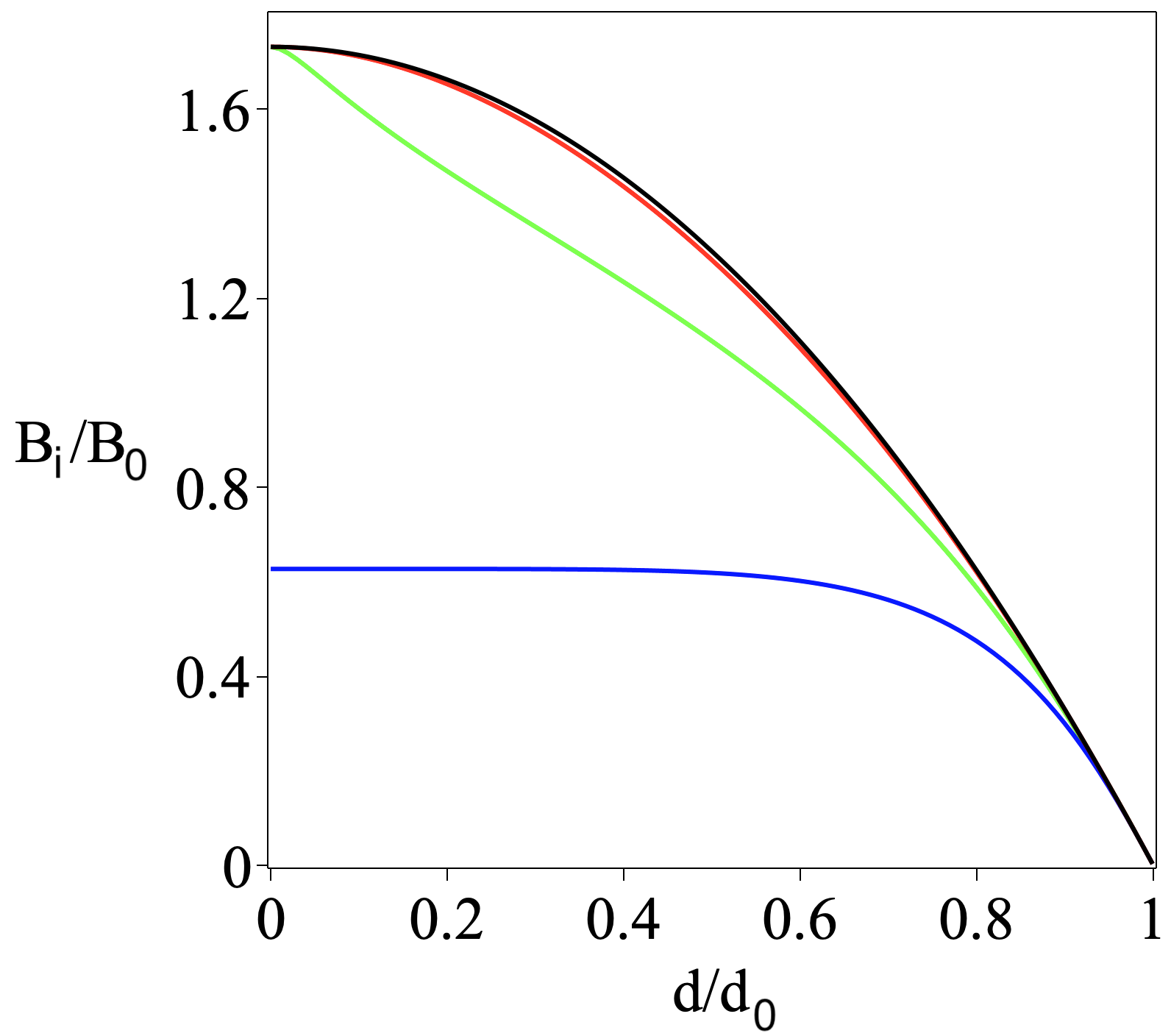}
\caption{(a) Fin current sheet length $L/d_0$ as a function of source separation $d/d_0$ in units of the flux interaction distance ($d_0=2F_0/(\pi B_0)$), where $\pm F_0$ are the fluxes of the sources and $B_0$ is the strength of the overlying field. The sheet length in the absence of reconnection is shown in black and for $C=\alpha/M_{A0}= 0.1$, 1, 5 in red, green and blue, respectively. (b) Inflow magnetic field strength ${B}_i/B_0$ for the same values of $C$.} 
\label{fig:L-of-d-with-rec}
\end{figure}

Next, in order to solve this equation for $L(d)$, we adopt a change of variable, setting $u=y/L$, which reduces it to
\begin{equation}\label{eq:fluxbalancenorec}
    L\left( 1 + \int_1^\infty 1-\frac{u(u^2-1)^{1/2}}{u^2+(d^2/L^2)} \id u\right) = \frac{\pi}{2}d_0, 
\end{equation}
the solution of which is surprisingly simple (see Appendix A), namely,
\begin{equation}\label{L-of-d-norec}
    L(d)=\sqrt{d_0^2-d^2},
\end{equation}
so that the current sheet length ($L$) vanishes at $d=d_0$ (as required), and it tends to  $d_0$ as $d$ approaches zero, as shown by the black curve in Fig. \ref{fig:L-of-d-with-rec}(a).  

%{\red David - please add the values C=0 and 0.1 and update the caption for this and the next figure.}
%
\begin{figure}
\centering
\includegraphics[width=0.4\textwidth]{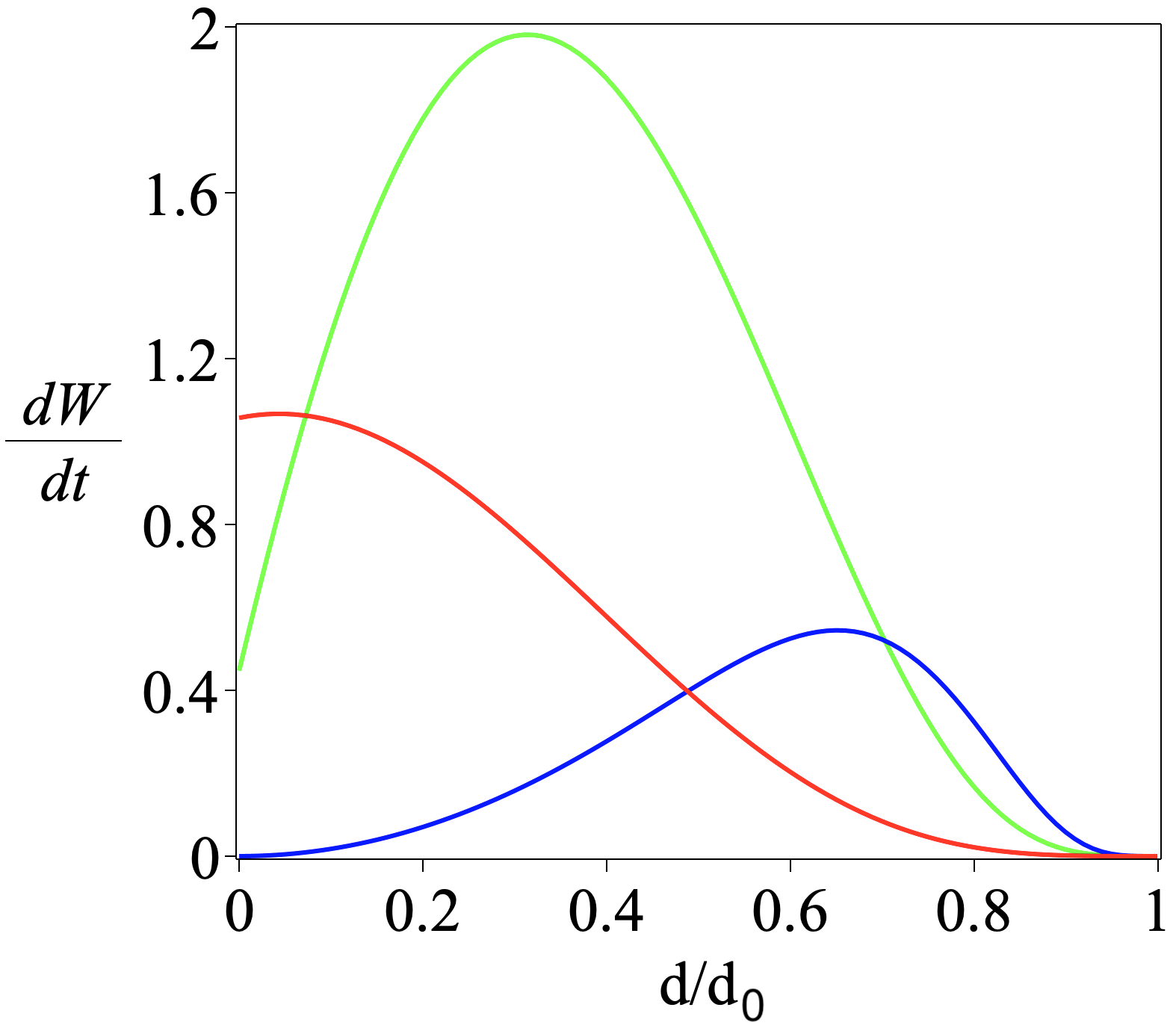}
%(b)\includegraphics[width=0.45\textwidth]{dWdt-of-d_plot_plusfloat.png}
\caption{Rate of thermal energy release as a function of source separation $d$   for  the fin current sheet model, with $C=\alpha/M_{A0}=0.1,$ 1 and 5 (red, green, and blue, respectively). }%(b) Shows  a comparison between the thermal energy release for the fin current sheet model (solid line) a floating current sheet (dashed line) for $C=100$ (a parameter choice for which the floating sheet model is valid for $0.99\gtrsim d\gtrsim 0.1$.} 
\label{fig:dWdt}
\end{figure}
\subsection{Fin Current Sheet Length with Reconnection}
\label{sec2.3}
If the plasma is not perfectly ideal, then, during the time that the flux patches approach one another, reconnection will commence in the current sheet. This will add flux above the current sheet, and the current sheet length will become smaller than when reconnection is prohibited. It  is then possible to determine the sheet length ($L$), the inflow speed ($v_i$) and the inflow magnetic field ($B_i$) in terms of the imposed values of the flux source separation ($2d$), the speed of approach ($v_0$) of the sources, and the overlying magnetic field strength ($B_0$), as follows.

The magnetic flux ($F_R$) added above the current sheet is equal to the net flux ($\psi$) into the sheet, namely,
\begin{equation}
    F_R=\int \frac{d\psi}{dt} \,\id t = \int v_{i}B_{i}\,\id t,
    \nonumber
\end{equation}
and, after defining the speed of approach of the flux sources as $v_0=\id(d)/\id t$, this becomes
\begin{align}
    F_R &= \int \frac{v_i}{v_0} B_i \id d.
    \label{eq:flux1}
\end{align}

In order to estimate the flux reconnected, we need to assume a typical inflow speed ($v_{i}=\alpha v_{Ai}$) to the reconnection site in terms of the local Alfv\'en speed ($v_{Ai}$), where $\alpha$ depends on the nature of the reconnection and the rate at which it is being driven, as described in, e.g., \cite{priest14}. Thus, reconnection may be slow, with $\alpha$ being tiny (say, $10^{-6}$) or it may be fast, with the maximum value of $\alpha$ for various fast models being in the range $0.01 - 0.1$ \citep[e.g.,][]{priest14,cassak17,pontin22}. Then the ratio of the inflow speed to the driving speed becomes
\begin{equation}
    \frac{v_{i}}{v_0} =\frac{\alpha v_{ Ai}}{v_0} = \frac{\alpha v_{{ A}0}}{v_0}\frac{B_{i}}{B_{0}}=\frac{\alpha}{M_{A0}} \frac{B_{i}}{B_{0}},
    \label{eq:vin}
\end{equation}
where $M_{A0}=v_0/v_{A0}$ and $v_{A0}=B_0/\sqrt{\mu \rho_{i}}$ is the Alfv\'en speed based on the overlying magnetic field ($B_0$) and the density ($\rho_{i}$) at the inflow to the current sheet. 
Thus, Equation (\ref{eq:flux1}) can be written as 
\begin{equation}
    F_R=C B_0 d_0 \int \frac{B_{i}^2}{B_0^2}\id d/d_0,
    \label{eq19}
\end{equation}
where $C=\alpha v_{A0}/v_0$ is a dimensionless constant. In this and in what follows, we for simplicity assume that $\alpha = constant$, but, if $\alpha$ is non-constant, the resulting energy release will lie between the different curves associated with $\alpha = constant$

\begin{figure*}
\centering
\includegraphics[width=0.7\textwidth]{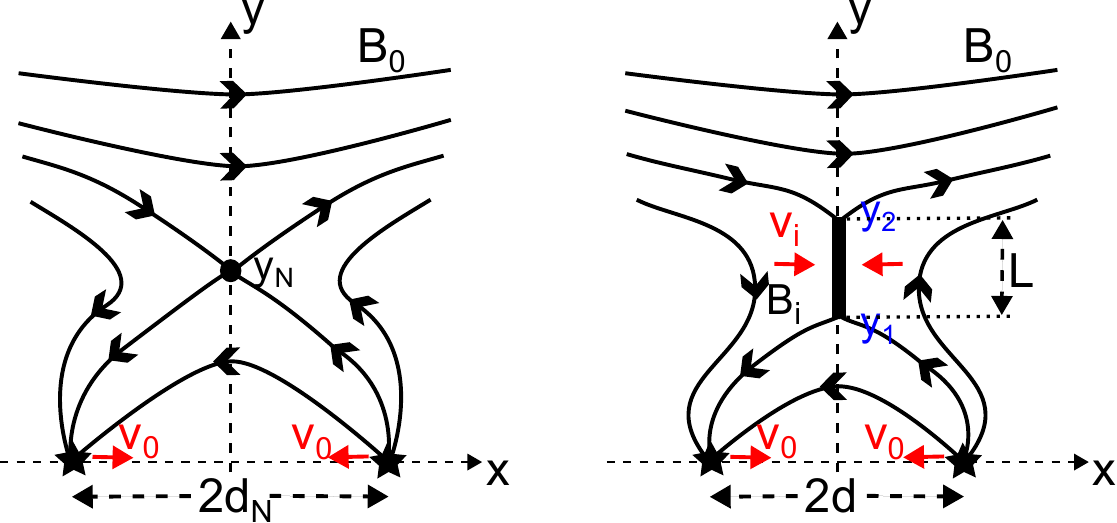}
\caption{The nomenclature for a floating current sheet. (a) An initial potential state with a null point at height $y_N<d_0$ due to two opposite polarity sources a distance $2d_N$ apart in a uniform horizontal field ($B_0$). (b) The field that is formed as the sources approach one another  with speeds $v_0$ to a separation distance of $2d<2d_N$, giving rise to a current sheet of length $L$ stretching from $y_1$ to $y_2$ on the $y$-axis, at which the inflow speed and magnetic field are $v_i$ and $B_i$, respectively.}
\label{fig4}
\end{figure*}
Hence, we modify Eqn.(\ref{eq:fluxbalancenorec}) to include the reconnected flux to give
\begin{equation}\label{eq:fluxbalancerec}
    L\left( 1 + \int_0^\infty 1-\frac{u(u^2-1)^{1/2}}{u^2+(d^2/L^2)} \id u \right) = \frac{\pi}{2}d_0-Cd_0 \int_{d_0}^d (B_i/B_0)^2\id  d/d_0.
\end{equation}
To estimate $B_{i}$ we use the value at the centre of sheet, by substituting $y=\half L$ into Eqn.(\ref{eq:by_2d}), giving
\begin{equation}
    B_{i} = B_0\frac{\sqrt{3}L^2}{L^2+d^2}.
    \label{eq:bin}
\end{equation}
Substituting into Eqn.(\ref{eq:fluxbalancerec}) and evaluating the integral on the left-hand side yields
\begin{equation}
    \frac{\pi}{2}\sqrt{L^2+d^2} = \frac{\pi}{2}d_0-C\int_d^{d_0} \frac{3L^4}{(L^2+d^2)^2}\id d,
\nonumber
\end{equation}
which we wish to solve for the unknown function $L(d)$.

To solve this equation for $L(d)$ we first differentiate to obtain
\begin{equation}\label{eq:L-of-d-DE}
    L\frac{\id L}{\id d}+d = C \frac{6L^4}{\pi(L^2+d^2)^{3/2}}.
\end{equation}
This can be integrated numerically to find $L(d)$ using the initial condition that $L(d_0)=0$. In practice we perform the integration from $d=d_0(1-\epsilon)$ with $\epsilon \ll 1$ since ${\id L}/{\id d}$ tends to $-\infty$ as $d$ approaches $d_0$. To find this initial condition we use the known solution without reconnection (Eq.~\ref{L-of-d-norec}) to approximate 
\begin{equation}
    L(d_0(1-\epsilon))=d_0\sqrt{1-(1-\epsilon)^2} \approx \sqrt{2\epsilon d_0}.
\end{equation}
Since $\alpha$ may vary between, say, 0.001 and 0.1, while $M_{A0}$ varies between, say, 0.001 and 0.01 for $v_0=1$ km s$^{-1}$ and $v_{A0}=10^3$ km s$^{-1}$, $C=\alpha/M_{A0}$ varies between 0.1 and 100. The current sheet length as a function of $d$ is plotted for selected values of $C$ in Figure \ref{fig:L-of-d-with-rec}(a). Expressions when $C$ is small or large are derived in Appendix B.

\subsection{Heating Rate}
\label{2.4}
The values of the inflow speed (Eq.~\ref{eq:vin}) and inflow magnetic field (Eq.~\ref{eq:bin}), together with the current sheet length (Fig.~\ref{fig:L-of-d-with-rec}), allow us to investigate how the rate of magnetic energy release depends on time and on the parameters.  

The rate at which magnetic energy is transported into the current sheet is given by the Poynting flux into the two sides of the sheet, and, if two fifths of this energy is converted into thermal energy in standing slow-mode shocks \citep{priest14}, then the  heating rate is
\begin{equation}
    \frac{\id {W}}{\id t} = \frac{4}{5}\frac{{v}_{i}{B}_{i}^2}{\mu_0}LL_S,
    \nonumber
\end{equation}
for our two-dimensional model, where $L_S$ is the extent of the current sheet out of the plane. In this expression, the inflow speed to the current sheet is $v_{i}=\alpha v_{Ai}=\alpha v_{A0}B_{i}/B_0$, and so the heating rate may be written
\begin{equation}
    \frac{\id W}{\id t} =  \frac{\id W_0}{\id t} \left(\frac{B_{i}}{B_0}\right)^3\frac{L}{d_0},
    \label{heatrate}
\end{equation}
where $d{W}_0/dt=0.8\alpha B_0^3d_0L_S/[\sqrt{\mu \rho_{i}} \mu]$. 
The resulting heating rate after substituting for $B_{i}^3$ and $L$ is shown in Fig. \ref{fig:dWdt} for several values of the parameter $C$.

\section{A Floating Current Sheet}\label{sec:float}
\subsection{General Treatment and Numerical Solution}
\label{sec{3.1}}

The so-called `floating current sheet' occurs when reconnection is initiated after a null point has formed in the corona, as studied previously by \cite{priest18} and \cite{syntelis19a}. 
%In two dimensions they used the same expression (\ref{heatrate}) for the heating rate, but with different expressions for the inflow magnetic field and sheet length, namely, from \cite{syntelis19a} \begin{equation}{B_{i}}=B_0\left(\frac{d_0}{d}-1\right)^{1/2}\frac{L}{d_0}.\label{eq28}\end{equation}and \begin{equation}L=\frac{d_0}{C^{1/2}(d_0/d-1)^{1/4}},\label{eq29}\end{equation} where all symbols have the same meanings as above.
\begin{figure}
\centering
(a)\includegraphics[height=0.24\textheight]{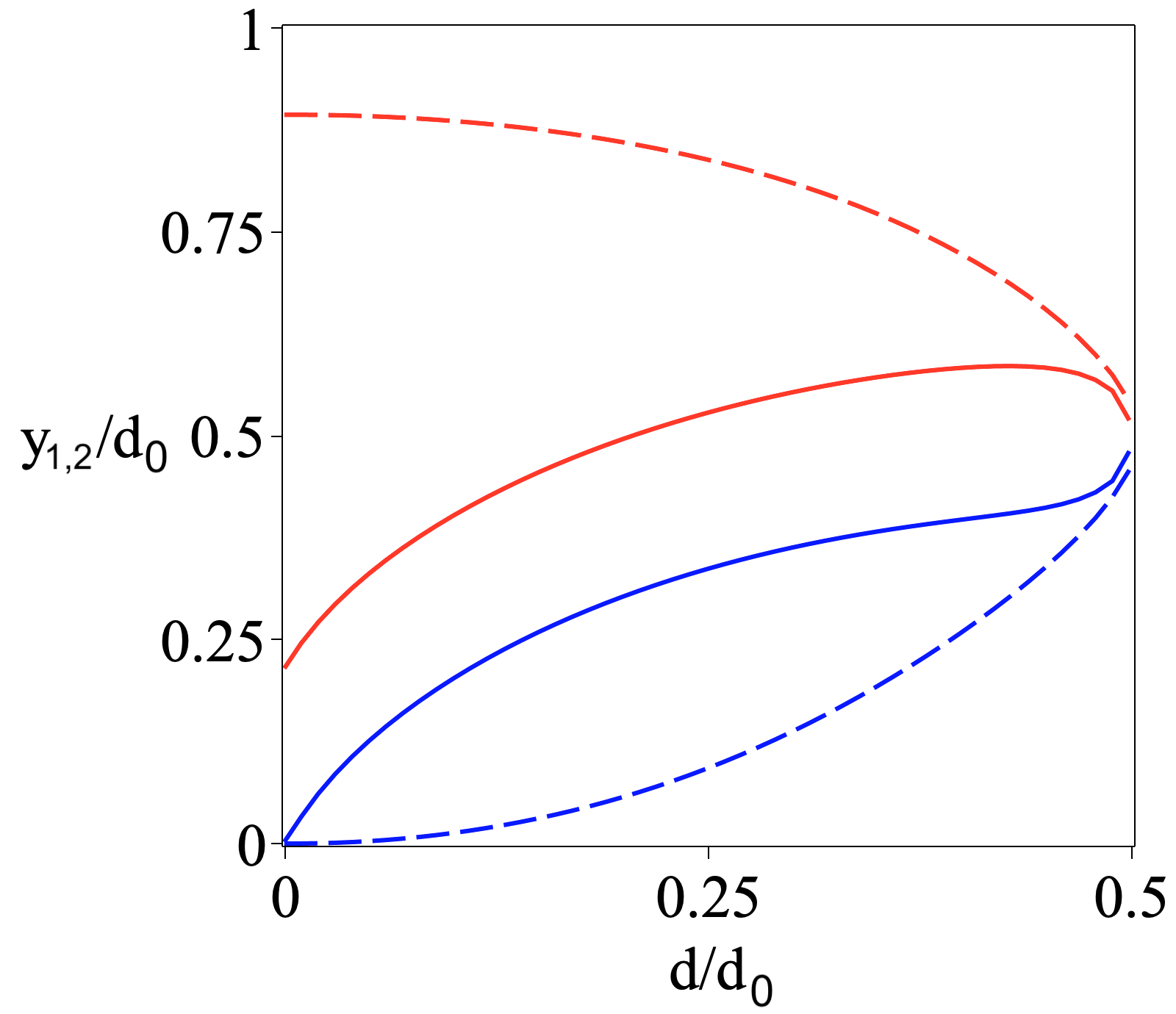}
(b)\includegraphics[height=0.24\textheight]{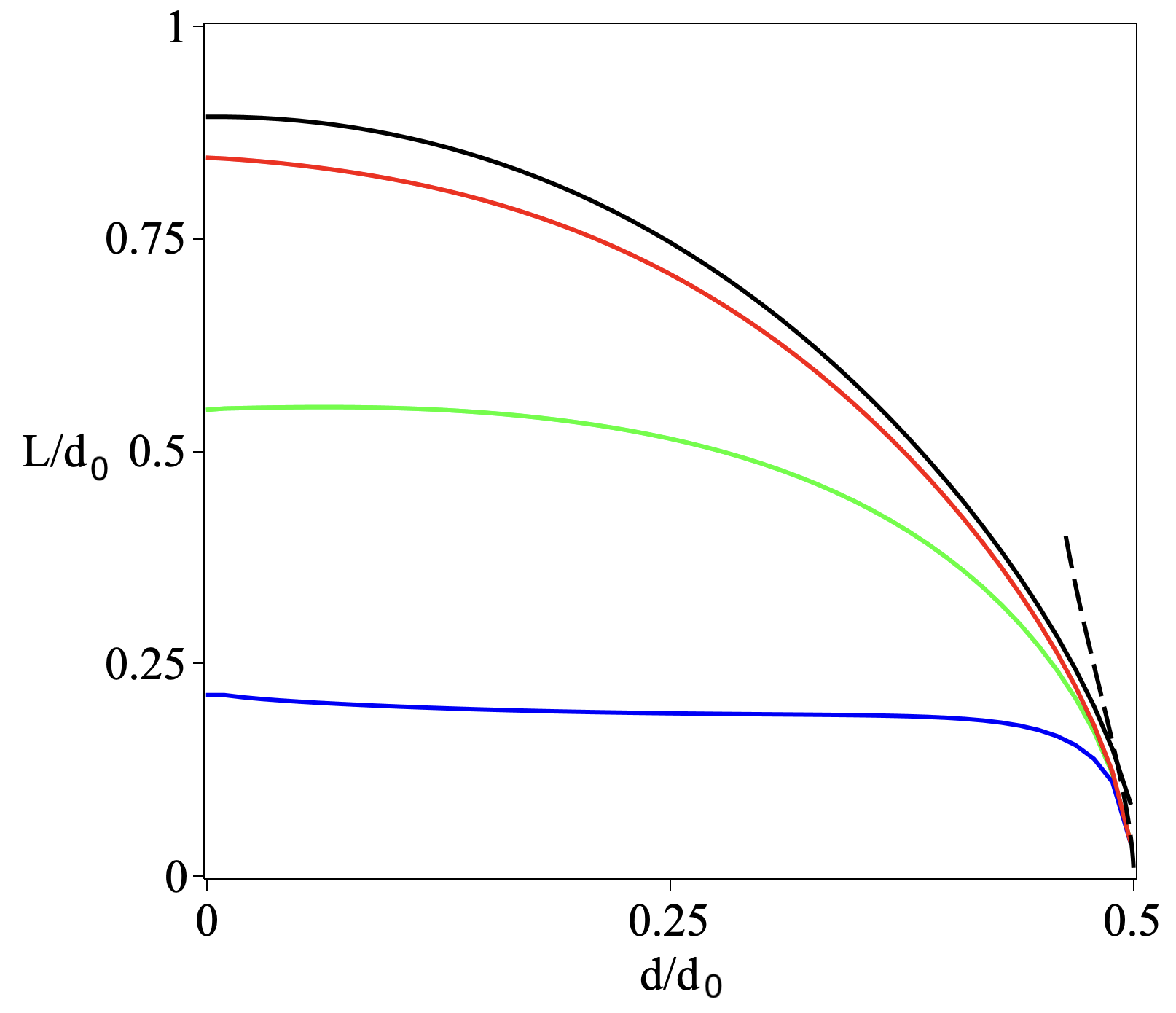}
(c)\includegraphics[height=0.24\textheight]{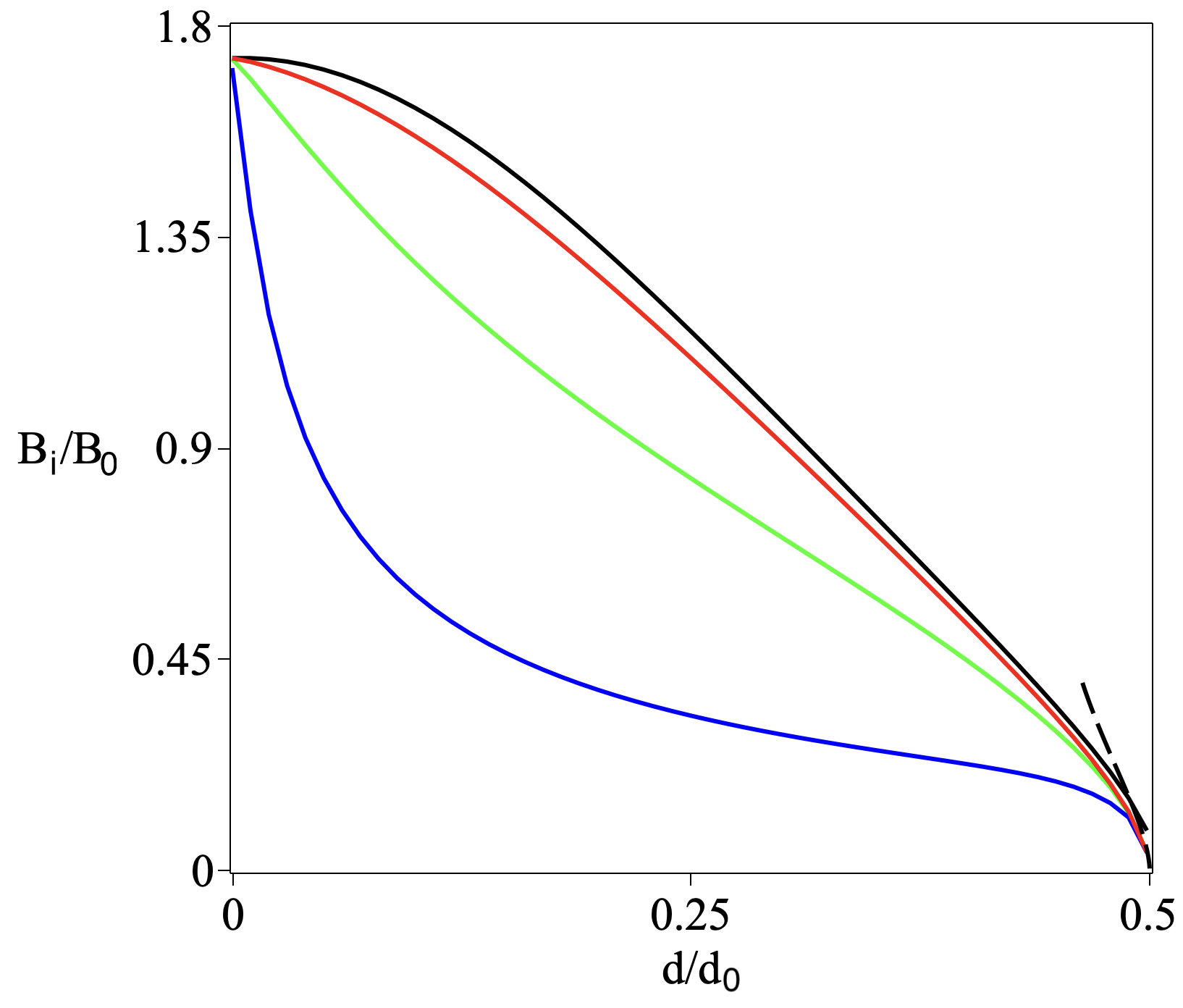}
\caption{Parameters of the current sheet for the floating sheet model when the initial source separation ($2d_N$) equals $d_0$, where $d_0$ is the flux interaction distance. (a) The endpoints $y_1$ (blue) and $y_2$ (red) of the current sheet in the absence of reconnection ($C=0$, dashed curves) and for $C=5$ (solid curves). (b) The length ($L$) of the floating current sheet ($L=y_2-y_1$), as well as (c) the magnetic field ($B_i$) at the input to the current sheet, for $C=0$ (black), $C=0.1$ (red), $C=1$ (green), and $C=5$ (blue). The approximations when $d$ is close to $d_N$ (for $C=0$) are shown as dashed curves in parts (b) and (c).} 
\label{fig5}
\end{figure}
\begin{figure}
\centering
(a)\includegraphics[width=0.45\textwidth]{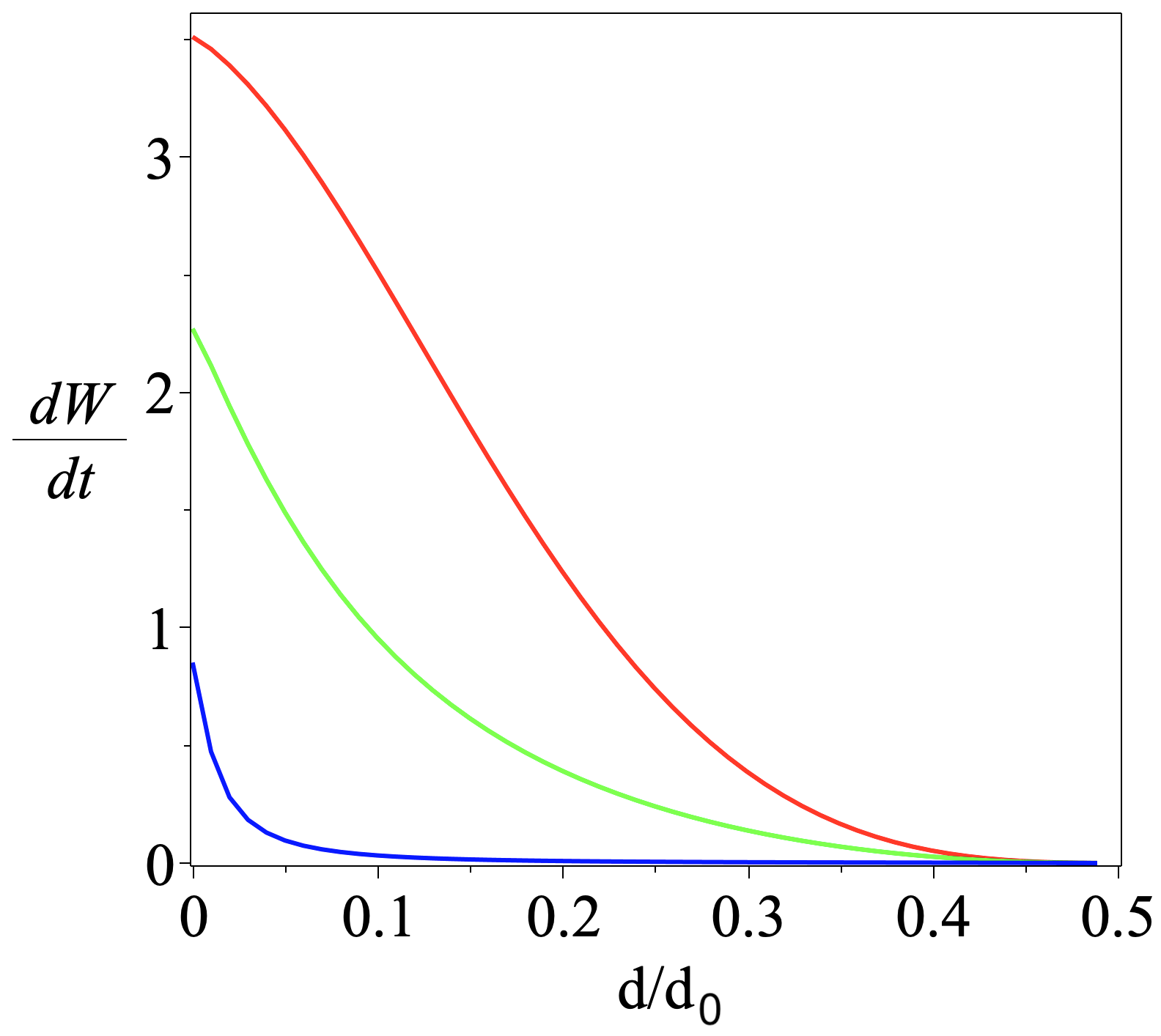}\\
(b)\includegraphics[width=0.45\textwidth]{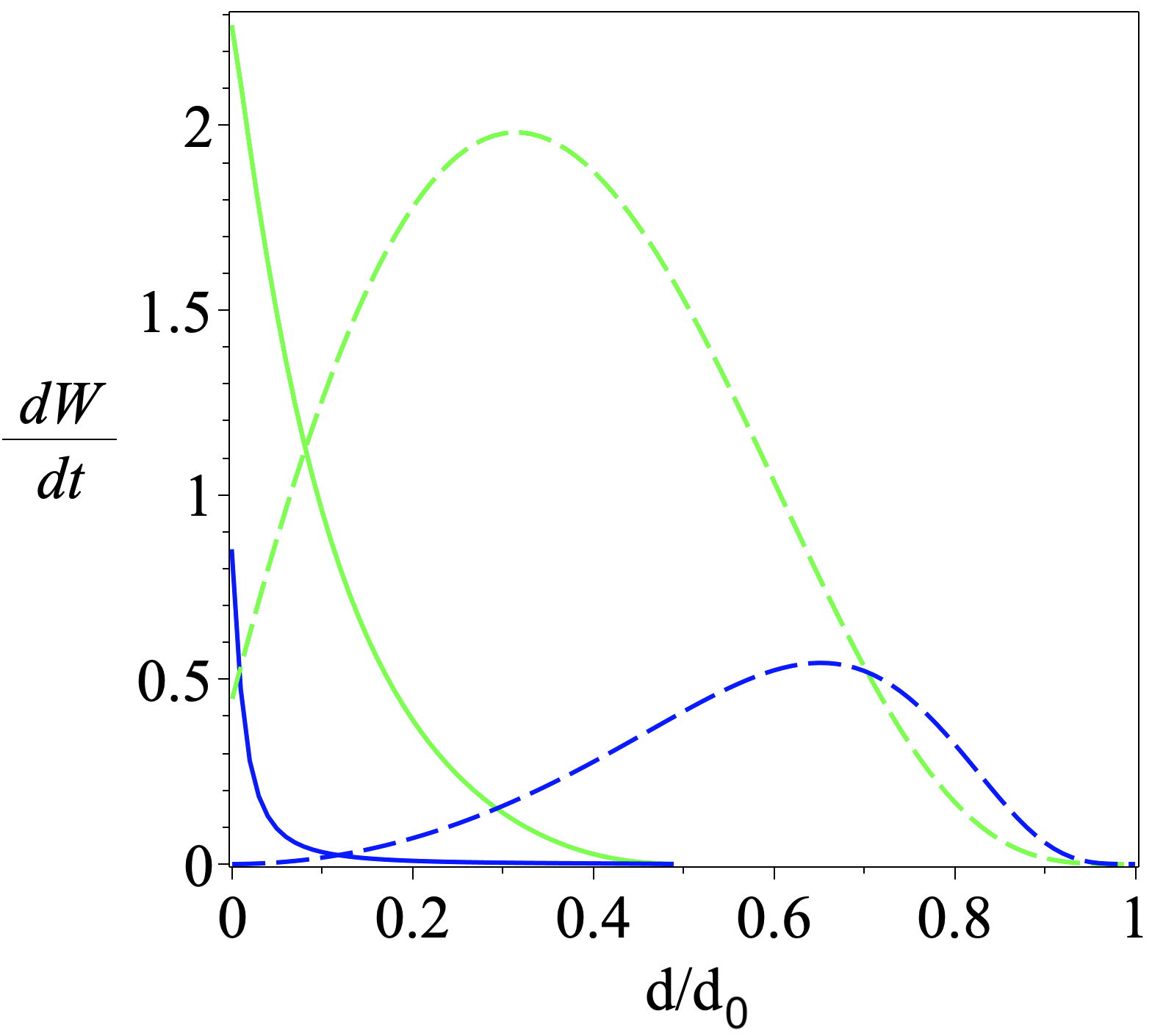}
\caption{(a) The rate of energy release for a floating current sheet with $d_N=\half d_0$ and $C=0.1, 1, 5$. (b) Comparison of the energy release rate for a fin current sheet (dashed) and floating current sheet (solid lines) for $C=1$ (green), and $C=5$ (blue).}
\label{fig6}
\end{figure}
However,
%in deriving these expressions, it was 
they assumed that $d$ is close to its initial value $d_N$, say, so that the magnetic field is close to potential and the current sheet is very short ($L\ll d$). In this section we therefore improve and generalise their analysis to long current sheets and nonpotential magnetic fields so as to be able to compare the results for a floating current sheet with those of a fin current sheet.

The improved technique employed here for calculating the three unknown variables, namely, the current sheet length ($L$) and the inflow speed ($v_i$) and field ($B_i$) to the current sheet, may be compared with the earlier technique for short current sheets adopted by \cite{syntelis19a} and \cite{priest18}. Here we simply use the two flux conditions above and below the sheet for $L$ and $B_i$ and then deduce $v_i$ from the reconnection rate ($v_i=\alpha v_{Ai}$, see Equation (\ref{eq:vin}) above). The earlier technique assumed as the first condition that $B_i=\half k L $, where $k$ comes from the form of the potential field near the null, and this gives the same result as Equation (\ref{eqBi}) below. The second condition was that the time rate of change of flux ejected below or above the sheet is equal to the flux ($v_iB_i$) into the current sheet, which relates $L$ to $v_i$. The final condition was the same as here, namely, $v_i=\alpha v_{Ai}$.

Suppose the magnetic field is in a potential state with no current sheet when the source separation is $d_N$ (Fig.\ref{fig4}a), so that its magnetic field in complex form is 
\begin{equation}
   B_x-iB_y  = B_0\left(\frac{z^2-d_N^2+d_0d_N}{z^2-d_N^2}\right),
   \label{eq16}
\end{equation}
where $z=x+iy$. This has a null point at $y=y_N$ on the $y$-axis, where
\begin{equation}
y_N^2=d_0d_N-d_N^2,
\nonumber
\end{equation}
and the magnetic field on the $y$-axis is
\begin{equation}
   B_x(0,y)  = B_0\left(1-\frac{d_0d_N}{y^2+d_N^2}\right),
   \nonumber
\end{equation}
so that the magnitude of the magnetic flux below the null point is 
\begin{equation}
   \psi_N=\int_0^{y_N}-B_x(0,y)\ dy  = B_0\left(-y_N+d_0\arctan \frac{y_N}{d_N}\right).
   \label{eq32}
\end{equation}

On the other hand, the total magnetic flux ($\psi_T$) across the $y$-axis is just
\begin{equation}
  \psi_T=\int_{0}^\infty B_0\ dy -F,
   \nonumber
\end{equation}
since a flux of amount F coming in from the left is diverted into the negative source.
The magnetic flux above the null point is therefore
\begin{equation}
  \int_{y_N}^\infty B_x(0,y)\ dy  = \psi_T+\psi_N= \int_0^\infty B_0\ dy-F+\psi_N,
   \nonumber
\end{equation}
since an extra flux of magnitude $\psi_N$ has been reconnected across the null point into the regions above and below it. Although this integral is infinite in view of the uniform field $B_0$ extending to infinity, it is useful to recast it as
\begin{equation}
  \int_{y_N}^\infty (B_x(0,y)-B_0)\ dy  = B_0y_N-F+\psi_N,
   \label{eq32a}
\end{equation}
where $F=\half \pi B_0 d_0$.

Then suppose the sources approach one another at speeds $v_0$, so that their separation is $2d$ and a current sheet of length $L$ forms, stretching along the $y$-axis from $y_1$ to $y_2$ (Fig.\ref{fig4}b). Then the magnetic field may be represented by 
\begin{equation}
    B_x-iB_y =B_0\frac{(z^2-z_1^2)^{1/2}(z^2-z_2^2)^{1/2}}{z^2-d^2},
\nonumber
\end{equation}
so that along the $y$-axis
\begin{align}
    B_y&=0,\ \ \ \ B_x(0,y)=-B_0\frac{(y_1^2-y^2)^{1/2}(y_2^2-y^2)^{1/2}}{y^2+d^2},     \label{eq34}
    \\  &{\rm when} \ \ y<y_1,\nonumber
\\
%\end{equation} \begin{equation}
    B_x&=0,\ \ \ \ B_y(0,y)=\pm B_0\frac{(y^2-y_1^2)^{1/2}(y_2^2-y^2)^{1/2}}{y^2+d^2},     \label{eq35}
    \\ &{\rm when} \ \ y_1<y<y_2,\nonumber
\\
%\end{equation} \begin{equation}
    B_y&=0,\ \ \ \ B_x(0,y)=B_0\frac{(y^2-y_1^2)^{1/2}(y^2-y_2^2)^{1/2}}{y^2+d^2},     \label{eq36}
    \\  &{\rm when} \ \ y>y_2.\nonumber
\end{align}
The current sheet endpoints ($y_1$ and $y_2$) and therefore the length $L$ of the current sheet are then given by the conservation of flux above and below the current sheet, namely,
\begin{equation}
\int_0^{y_1}-B_x(0,y)\ dy=\int_{y_2}^\infty (B_x-B_0)\ dy+\half \pi B_0d_0-B_0y_2=\psi_N+F_R,
\label{eq37}
\end{equation}
where an argument similar to deriving Eqn.(\ref{eq32a}) has been used. Here $B_x(0,y)$ is given by Eqns.(\ref{eq34}), (\ref{eq36}), $\psi_N$ by Eqn.(\ref{eq32}), and the reconnected flux ($F_R$) by an equation similar to Eqn.(\ref{eq19}), namely,
\begin{equation}
    F_R=CB_0 d_0 \int_{d_N}^d B_i^2\ \id d,
    \label{eq38}
\end{equation}
with $C=\alpha v_{A0}/v_0$.
In this expression for $F_R$, the function $B_i(d)$, namely, the field just outside the midpoint ($y_N$) of the current sheet is given by the value of $B_y$ from Eqn.(\ref{eq35}) evaluated at $y_N=\half (y_1+y_2)$, namely, 
\begin{equation}
B_i=B_0\frac{(y_2-y_1)(y_2+3y_1)^{1/2}(3y_2+y_1)^{1/2}}{(y_1+y_2)^2+4d^2}.
\label{eq39}
\end{equation}
Thus,   $B_i$ and $L$ are given by the coupled equations (\ref{eq37}), (\ref{eq38}), and (\ref{eq39}). The resulting solutions  can be obtained numerically as follows. When $d=d_0$, $y_1=y_2=y_N$ and $F_R=0$. Starting from these values, $d$ is decreased by small increments from $d=d_0$ to $d=0$. For each value of $d$ the pair of equations in Eq.~(\ref{eq37}) is solved for $y_1$ and $y_2$ using the \texttt{fsolve} command in \texttt{Maple}\textsuperscript{TM} \footnote{Maple is a trademark of Waterloo Maple Inc.}. These values are then used to find $B_i$ from (\ref{eq39}) and then to update $F_R$ from (\ref{eq38}), following which $d$ is decreased by a small increment and the process repeated. The numerical solutions are plotted in Fig.\ref{fig5} for $d_N=0.5 d_0$ and $C=0$, 0.1, 1 and 5.  In the limit as the current sheet length becomes small compared with the flux interaction distance ($L \ll d_0$), 
%similar to those in Eqns.(\ref{eq38}) and (\ref{eq39}) found by \cite{syntelis19a}, as proved 
%Appendix B shows 
we show in the following section that these reduce to 
\begin{equation}
{B_{i}}=B_0\left(\frac{d_0}{d_N}-1\right)^{1/2}\frac{L}{d_0},
\nonumber
\end{equation}
as in \cite{syntelis19a}, and
\begin{equation}
\frac{L^2}{d_0^2}\log_e\frac{d_0}{L}=\frac{f(d_N/d_0)}{(d_N/d_0-1)^{1/2}}\frac{D}{d_0},
\nonumber
\end{equation}
where
\begin{equation}
f\left(\frac{d_N}{d_0}\right)= \left( \frac{d_0}{d_N}-2\right) \arctan \sqrt{\frac{d_0}{d_N}-1}
    +\sqrt{\frac{d_0}{d_N}-1}.
\nonumber
\end{equation}

The rate of energy release is given by the same expression as for the fin current sheet, namely, Eqn.(\ref{heatrate}), but with $B_i$ and $L$ now given by Eqns.(\ref{eq37})--(\ref{eq39}). It is plotted in Fig.\ref{fig6}a for $C=0.1$, 1 and 5.

\subsection{The Current Sheet Length and Inflow Magnetic Field when the Sheet is Short}
\label{sec3.2}

Consider the floating current sheet illustrated in Fig.\ref{fig4}. Initially, the source separation is $2d_N$ and the magnetic field (\ref{eq16}) is
\begin{equation}
B_x-iB_y=B_0\frac{z^2+y_N^2}{z^2-d_N^2},
\nonumber
\end{equation}
which is potential and possesses a null point at height $y_N$ given by
\begin{equation}
y_N^2=d_0d_N-d_N^2.
\nonumber
\end{equation}
The magnetic flux (\ref{eq32}) below the null point can be written
\begin{equation}
\psi_N=-B_0\int_0^{y_N} \frac{y_N^2-y^2}{y^2+d_N^2} dy.
\nonumber
\end{equation}

When the sources approach one another to a separation $2d<2d_N$, a vertical current sheet forms stretching from $z_1$ to $z_2$ on the $y$-axis (Fig.~\ref{fig4}b), and the magnetic field becomes
\begin{equation}
B_x-iB_y=B_0\frac{(z^2-z_1^2)^{1/2}(z^2-z_2^2)^{1/2} }{z^2-d^2}.
\nonumber
\end{equation}
The resulting flux below the current sheet can then be written
\begin{equation}
\psi_S=-B_0\int_0^{y_1} \frac{(y_1^2-y^2)^{1/2}(y_2^2-y^2)^{1/2}}{y^2+d^2} dy +F_R,
\nonumber
\end{equation}
where $F_R$ is the reconnected flux.

The value of the current sheet length when $d$ is close to $d_N$, so that the sheet length ($L$) is much smaller than $d_N$, is given by equating the two fluxes $\psi_N$ and $\psi_S$ and linearising by writing $y_1=y_N-Y_1$, $y_2=y_N+Y_2$, and $d=d_N-D$, where $Y_1\ll y_N$, $Y_2\ll y_N$, and $D\ll d_N$. Then, equating $\psi_S$ and $\psi_N$ and omitting $F_R$ since it is of higher order in the small quantities than those we shall ultimately keep,
\begin{equation}
\int_0^{y_1} \frac{(y_1^2-y^2)^{1/2}(y_2^2-y^2)^{1/2}}{y^2+d^2} dy= \int_0^{y_N} \frac{y_N^2-y^2}{y^2+d_N^2} dy.
\nonumber
\end{equation}

After subtracting the integral of the initial flux up to $y=y_1$ from both sides, this may be rewritten
\begin{equation}
I_B = I_C+I_A,
\nonumber
\end{equation}
where
\begin{equation}
I_B =
\int_0^{y_1}  \frac{(y_1^2-y^2)^{1/2}(y_2^2-y^2)^{1/2}-(y_N^2-y^2)}{y^2+d^2} dy,
\nonumber
\end{equation}
\begin{equation}
I_C =
\int_0^{y_N}  \frac{y_N^2-y^2}{y^2+d_N^2}-\frac{y_N^2-y^2}{y^2+d^2} dy,
\nonumber
\end{equation}
which are evaluated in Appendix C,
and
\begin{equation}
I_A =
\int_{y_1}^{y_N}  \frac{y_N^2-y^2}{y^2+d^2} dy.
\nonumber
\end{equation}

Each of these integrals is small, and, since $I_B \sim  L^2y_N\log_e L/(d_0d_N)$, $I_C \sim D$, and $I_A \sim L^2 y_N/(d_0d_N)$, we shall to lowest order neglect $I_A$ compared with the other two integrals.
Thus, equating the expressions (\ref{IB}) for $I_B$ and (\ref{IC}) for $I_C$ in Appendix C, we obtain the result that 
\begin{equation}
\frac{L^2}{d_0^2}\log_e\frac{d_0}{L}=\frac{4f(d_N/d_0)}{(d_0/d_N-1)^{1/2}}\frac{D}{d_0},
\end{equation}
so that the current sheet length ($L$) in the combination $L^2 \log_e L$ is proportional to the small displacement ($D$) of the sources from $d_N$, where $f(d_N/d_0)$ is given by Eq.(\ref{f}). When $d_N$ is close to $d_0$, this is well behaved, since the expression for $L$ behaves like
\begin{equation}
\frac{L^2}{d_0^2}\log_e\frac{d_0}{L}=\frac{4(d_0/d_N-1)^2}{3}\frac{D}{d_0},
\nonumber
\end{equation}
which is plotted as the dashed curve in Fig. \ref{fig5}b.

Furthermore, Equation (\ref{eq39}) for the magnetic field at the inflow to the current sheet, namely, 
\begin{equation}
B_i=B_0\frac{(y_2-y_1)(y_2+3y_1)^{1/2}(3y_2+y_1)^{1/2}}{(y_1+y_2)^2+4d^2}.
\nonumber
\end{equation}
may be approximated to lowest order for short current sheets when $L \ll d_0$, by putting $y_2-y_1\approx L$, $y_2+3y_1 \approx 3y_2+y_1 \approx 4y_N$, and $y_1+y_2\approx 2y_N$, so that
\begin{equation}
B_i=B_0\frac{Ly_N}{y_N^2+d_N^2},
\nonumber
\end{equation}
where $y_N^2=d_0d_N-d_N^2$, so that we recover the same expression as in \cite{syntelis19a}, namely,
\begin{equation}
B_i=B_0\left(\frac{d_0}{d_N}-1\right)^{1/2}\frac{L}{d_0},
\label{eqBi}
\end{equation}
which is plotted as the dashed curve in Fig. \ref{fig5}c.

%%%%%%%%%%
\section{Conclusion}\label{sec:conc}
Observations from the Sunrise balloon mission \citep{solanki17a,smitha17} and more recently from the Solar Orbiter and Solar Probe missions have raised the possibility that  magnetic reconnection driven by photospheric flux cancellation may be providing an important contribution to the heating of the 
%solar corona 
 Sun's atmosphere and the acceleration of the solar wind, as suggested by \cite{pontin24} building on previous proposals \citep{peter19,chitta20,chen21,tripathi21,panesar21,raouafi23}.

 The main implication of the present work is that, when oppositely directed fragments of photospheric magnetic flux approach one another, in the build-up to flux cancellation they may drive the formation of a current sheet between them, and so give rise to the heating and jets of plasma that could contribute to heating the chromosphere and corona and  accelerating the solar wind.  Previous analyses of energy release by such flux cancellation \citep{priest18,priest21a,syntelis19a,syntelis21} have demonstrated that, based on newly observed flux cancellation rates, the process can provide a significant contribution to heating the chromosphere and corona. Here we showed that cancellation at a fin current sheet -- consistent with observations of energy release in the low atmosphere -- is likely to be even more effective and produce much more heating. Furthermore, since the heating scales roughly as the square of the sheet length, dropping the assumption of a small current sheet also leads to much more heating in a floating current sheet than before. The net result of this theory is to put the suggestion of coronal heating by reconnection driven by flux cancellation on a much firmer foundation. In future, it will be interesting to compare with simulations designed to analyse the process in more detail.

The previous theory for reconnection by flux cancellation \citep{priest18,priest21a,syntelis19a,syntelis21} made two assumptions that we improve upon here. The first was that the current sheet develops about a coronal null point, to give a so-called ``floating current sheet".
However, whereas the flux from opposite-polarity flux fragments will not be connected when they are so far apart that their separation distance exceeds the flux interaction distance ($d>d_0$), as soon as it becomes equal ($d=d_0$) and decreases further ($d<d_0$) a null point (or separator) will form in the solar surface and a vertical current sheet will develop and grow.  This type of current sheet we term a ``fin sheet" and analyse its growth for the first time here. If there is no reconnection, it will grow in length as the flux sources approach, up to a maximum length of $L=d_0$ as their separation tends to zero. At the same time, the magnetic field at the sheet increases from zero up to a maximum value of $\sqrt 3B_0$, where $B_0$ is the overlying field strength. 

When reconnection takes place, the sheet length and inflow magnetic field are smaller, as shown in Fig.~\ref{fig:L-of-d-with-rec}, while the rate of energy release varies in a way shown in Fig.~\ref{fig:dWdt} and is usually larger than the rate from a floating current sheet.  Examining Fig.~\ref{fig6}(b) we see that for these parameters the rate of energy release for the fin current sheet remains larger than the floating sheet until the flux patches are very close to one another (while $d/d_0\gtrsim 0.1$). Integrating the area under the curves in Fig. \ref{fig6}(b) (and dividing by $v_0=\id d/\id t$) allows the total energy released during the interaction to be quantified. For $C=1$ the fin current sheet releases approximately a factor of 5 more energy ($105.0/v_0$ compared with $26.2/v_0$ in non-dimensional units), while for $C=5$ the difference in nearly a factor of 10 ($23.67/v_0$ compared with $2.52/v_0$).

The second assumption of previous theory that we focus on here is that a floating current sheet possesses a very small length (much smaller than its altitude). 
In Section \ref{sec:float} we dropped this assumption and allowed the sheet length to grow  to a substantial fraction of the flux interaction distance ($d_0$), the actual value depending on the rate of reconnection, as shown in Fig. \ref{fig5}b. Meanwhile, the inflow field strength also varies and becomes of order the overlying field strength as the separation between the flux sources decreases (Fig. \ref{fig5}c). At the same time, the rate of energy release grows as the flux sources approach one another, reaching a maximum value at the moment of flux cancellation (Fig. \ref{fig6}a).

One argument for the existence of  a floating current sheet is  that the approach of nearby photospheric flux fragments of opposite polarity may be an intermittent process, so that the reconnection could start and stop, allowing the current sheet to dissipate completely before restarting as the null point (or separator) rises in the atmosphere. Another argument is that such sheets will naturally form when new motions of photospheric flux sources occur in fields that are already topologically complex with null points or separators present.

Here on grounds of simplicity we have discussed current sheet formation in simple, idealised, two-dimensional magnetic fields affected by simple boundary flows, but the results are highly informative for the likely behaviour of  topologically much more complex solar magnetic fields of a non-potential three-dimensional nature.  First of all, current sheets may also form and be the location of magnetic reconnection and energy release in non-potential magnetic fields \citep[e.g.,][]{longcope05b,priest14,pontin22}.  Furthermore, they may form around both null points and separators in three-dimensional magnetic fields of complex topology, as well as around quasi-nulls and quasi-separators in magnetic fields of simple topology (i.e., with no nulls or separators) but complex geometry \citep[e.g.,][]{longcope05b,priest14,pontin22}. In addition, in a complex configuration with complex boundary flows, the resulting current sheets may have arbitrary inclination and be curved.  Analysing such complex situations in future by both analytical and computational means will be a challenge but will hopefully be helped by the understanding developed here.

%Comparisons are shown in Figure \ref{fig:dWdt}(b). Note that these expressions for $\bar{B}_i$ and $\bar{L}$ are invalid when $d$ is too close to $d=0$ and $d=1$. Thus, for example, near $d=0$, $\bar{B}_i\approx d^{-1/4}C^{-1/2}$, and so, if $\bar{B}_i<1$, we need $d>C^{-2}$. Also, near $d=1$ where $d=1-\epsilon$, say, if $\bar{L}<1$, we need $\epsilon>C^{-2}$. Thus, for example, when $C=10$, we need $d$ to be further from $0$ and $1$ by 0.01.
% \begin{figure}
% \centering
% \includegraphics[width=0.45\textwidth]{fig4.png}
% \caption{Comparison of the rate of energy release as a function of source separation $d$   for $C=\alpha/M_{A0}=100$ between the cases of a fin current sheet and a floating current sheet.} 
% \label{fig4}
% \end{figure}

\section*{Acknowledgements}
The authors thank the anonymous referee for constructive suggestions which have helped to improve the paper.
D.P.~gratefully acknowledges support through an Australian Research Council Discovery
Project (DP210100709).

%%%%%%%%%%%%%%%%%%%%%%%%%%%%%%%%%%%%%%%%%%%%%%%%%%
\section*{Data Availability}

No new data were generated or analysed in support of this research.

%%%%%%%%%%%%%%%%%%%% REFERENCES %%%%%%%%%%%%%%%%%%

% The best way to enter references is to use BibTeX:

%\bibliographystyle{mnras}
%\bibliography{fin} 

% Alternatively you could enter them by hand, like this:
% This method is tedious and prone to error if you have lots of references
%\begin{thebibliography}{99}
%\bibitem[\protect\citeauthoryear{Author}{2012}]{Author2012}
%Author A.~N., 2013, Journal of Improbable Astronomy, 1, 1
%\bibitem[\protect\citeauthoryear{Others}{2013}]{Others2013}
%Others S., 2012, Journal of Interesting Stuff, 17, 198
%\end{thebibliography}

%%%%%%%%%%%%%%%%%%%%%%%%%%%%%%%%%%%%%%%%%%%%%%%%%%

\appendix
\section{The solution for $L(d)$ for the fin current sheet}
{\bf  
Eqn.(\ref{eq:fluxbalancenorec}) is of the form
\begin{equation}
L(1+I)=\frac{\pi}{2} d_0,
\label{int}
\end{equation}
where
\begin{equation}
I=\int_1^\infty \left(1-\frac{u\sqrt{u^2-1}}{u^2+c}\right)du
\nonumber
\end{equation}
and $c=d^2/L^2$.
Changing the variable from $u$ to $U$, where $u^2=U^2+1$, we find
\begin{equation}
I=\int_1^\infty  du -\int_0^\infty \frac{U^2}{U^2+c+1}dU
\nonumber
\end{equation}
or 
\begin{equation}
I=\int_1^\infty  du -\int_0^\infty 1-\frac{c+1}{U^2+c+1}dU
\nonumber
\end{equation}
or
\begin{equation}
I=\left[\frac{\arctan({U/\sqrt{c+1}})}{\sqrt{c+1}}\right]_0^\infty-1=\frac{\pi}{2}\sqrt{c+1}-1
\nonumber
\end{equation}
Thus, Equation (\ref{int}) becomes
\begin{equation}
L\frac{\pi}{2}\sqrt{\frac{d^2}{L^2}+1}=\frac{\pi}{2}d_0,
\nonumber\end{equation}
or
\begin{equation}
L^2=d_0^2-d^2,
\nonumber\end{equation}
namely, Eqn.(\ref{L-of-d-norec}), as required.
}

\section{Evaluation of Fin Sheet Length ($L$) when $C$ Small or Large}

The fin sheet length ($L$) is given by Eqn.(\ref{eq:L-of-d-DE}). When $C=\alpha/M_{A0}$ is so small  that $6C/\pi=\epsilon_1\ll 1$, the solution for $L$ is close to $L_0=(d_0-d)^{1/2}$, and so may be written $L=L_0+\epsilon_1L_1$, where
\begin{align*}
L_1=&\frac{15d_0}{48(d_0^2-d^2)^{1/2}}(\arcsin (d/d_0)-\pi/2)
\\&+\frac{1}{48}(8(d/d_0)^5-26(d/d_0)^3+33(d/d_0)-15).
\nonumber
\end{align*}
As $d$ tends to zero, we see that $L_1$ tends to $-5( \pi/2 +1)/16$.

On the other hand, when $C$ is so large that $\epsilon_2=\pi/(6C)\ll 1$, there is a boundary layer  near $d=d_0$ of width $\delta = (1/2) \epsilon_2^{1/2}d_0$, such that $L = \epsilon_2^{1/4} d$ when $0<d < (1-\delta)d_0$ and $L=(d_0^2-d_0d)^{1/2}$ when $1-\delta < d/d_0 <1$.
The result of numerically integrating Eqn.(\ref{eq:L-of-d-DE}) with this initial condition for different values of the constant $C$ is shown in Fig. \ref{fig:L-of-d-with-rec}(a) with the corresponding values of the inflow field $B_i$ to the current sheet (Eqn.(\ref{eq:bin})) in Fig. \ref{fig:L-of-d-with-rec}(b).

\section{Evaluation of the Integrals $I_B$ and $I_C$}

The integral $I_B$ may be evaluated by assuming $Y_2=Y_1$ and changing the variable from $y$ to $Y$, where $y_N-y=\half LY$, so that it becomes
\begin{equation}
I_B \approx {\half L}\int_1^{2y_N/L}\frac{(Y^2-1)^{1/2}(4y_N/L-Y)+Y^2-4y_NY/L}{4(y_N^2+d_N^2)/L^2-4y_N/L+Y^2}dY,
    \nonumber
\end{equation}
which is dominated by values near the upper limit ($2y_N/L \gg 1$), where $(Y^2-1)^{1/2} \approx Y-1/(2Y)$ and the denominator approximates to $4(y_N^2+d_N^2)/L^2=4d_0d_N/L^2$. Thus the integral may be approximated by
\begin{equation}
I_B \approx \frac{L^3}{8d_0d_N}\int_1^{2y_N/L}\left(Y-\frac{1}{2Y}\right)\left(\frac{4y_N}{L}-Y\right)+Y^2-\frac{4y_NY}{L}dY,
    \nonumber
\end{equation}
or
\begin{equation}
I_B \approx -\frac{y_NL^2}{4d_0d_N}\int_1^{2y_N/L}\frac{1}{Y}dY=-\frac{y_N^2}{4d_0d_N} \log_e\frac{2y_N}{L},
    \nonumber
\end{equation}
where $y_N/d_N=(d_0/d_N-1)^{1/2}$, so that
\begin{equation}
    I_B \approx -\left(\frac{d_0}{d_N}-1\right)^{1/2}\frac{L^2}{d_0^2}\log_e \frac{d_0}{L}.
\label{IB}
\end{equation}

Next, the integral $I_C$ may be written
\begin{equation}
I_C =
\int_0^{y_N}  \frac{y_N^2-y^2}{y^2+d_N^2}\left(\frac{y_N^2+d_N^2}{y^2+d_N^2-2d_ND}-1 \right) dy,
\nonumber
\end{equation}
or
\begin{equation}
I_C \approx
\int_0^{y_N}  \frac{2d_ND(y_N^2-y^2)}{(y^2+d_N^2)^2}  dy,
\nonumber
\end{equation}
where
\begin{equation}
\int  \frac{1}{(x^2+c^2)^2}  dx = \frac{1}{2c^3} \arctan \frac{x}{c} + \frac{x}{2c^2(x^2+c^2)},
\nonumber
\end{equation}
and 
\begin{equation}
\int  \frac{x^2}{(x^2+c^2)^2}  dx = \frac{1}{2c} \arctan \frac{x}{2c} - \frac{x}{2c(x^2+c^2)},
\nonumber
\end{equation}
and so, after some manipulation, our integral becomes
\begin{equation}
    I_C=-D f\left(\frac{d_N}{d_0}\right), 
    \label{IC}
\end{equation}
where
\begin{equation}
f\left(\frac{d_N}{d_0}\right)= \left( \frac{d_0}{d_N}-2\right) \arctan \sqrt{\frac{d_0}{d_N}-1}
    +\sqrt{\frac{d_0}{d_N}-1}.
\label{f}
\end{equation}

%%%%%%%%%%%%%%%%%%%%%%%%%%%%%%%%%%%%%%%%%%%%%%%%%%

% Don't change these lines
\bsp	% typesetting comment
\label{lastpage}
\end{document}